# Exploring Selfish Trends of Malicious Mobile Devices in MANET

Dr. P.K.Suri and Kavita Taneja

**Abstract**—The research effort on mobile computing has focused mainly on routing and usually assumes that all mobile devices (MDs) are cooperative. These assumptions hold on military or search and rescue operations, where all hosts are from the same authority and their users have common goals. The application of mobile ad hoc networks (MANETs) as open networks has emerged recently but proliferated exponentially. Energy is a valuable commodity in MANETs due to the limited battery of the portable devices. Batteries typically cannot be replaced in MANETs, making their lifetime limited. Diverse users, with unlike goals, share the resources of their devices and ensuring global connectivity comes very low in their priority. This sort of communities can already be found in wired networks, namely on peer-to-peer networks. In this scenario, open MANETs will likely resemble social environments. A group of persons can provide benefits to each of its members as long as everyone provides his contribution. For our particular case, each element of a MANET will be called to forward messages and to participate on routing protocols. A selfish behavior threatens the entire community and also this behavior is infectious as, other MDs may also start to perform in the same way. In the extreme, this can take to the complete sabotage of the network. This paper investigates the prevalent malicious attacks in MANET and analyzes recent selfish trends in MANET. We analyzed the respective strengths and vulnerabilities of the existing selfish behaviour prevention scheme.

——————————— ◆ ———————————

## 1 INTRODUCTION

Since the first form of wireless ad hoc networks in the DARPA packet radio networks in the 1970s, it becomes an interesting research entity in the computer industry. During the last couple of years remarkable improvements are made in the research of ad hoc networks. Due to its possibility to fashion and manage a network without any central management, MANET is characterized as the art of networking without a network [1]. The success of MANETs is however perilously tied to their capability of transporting a wide gamut of applications with varying Quality of Service (QoS) requirements and providing service despite failures in the underlying network. Emerging applications from conventional to the social or commercialized ad hoc networks are bringing new, less explored arena of challenges. Non cooperation among MDs is one of these and it is a serious threat to the very existence of such networks. This kind of problem was already noticed on peer-to-peer file distribution systems like in Gnutella [2], the number of sites providing valuable content to the network was a small part of the number of users retrieving information. MANETs by their very nature are stochastic – in part because of the random environmental conditions and mobility of the MANET nodes, and in part because of the scarcity and variability of resources. Energy conservation is a critical issue in MANETs, which have nodes powered by batteries only. Thus, the lust to preserve energy, forces MDs to restrict use of their own resources to carry traffic of others. Such MD is undermined by selfish behavior, where its actions increases benefit for that MD while decreasing the average benefit for all MDs. Current schemes of detecting MD misbehavior in MANET are mostly centered on using incentives [3], reputation [4] or Game-based mechanisms [5] to achieve the desired effect of MDs cooperation. In an adhoc network one of the major concerns is how to increase the routing security in presence of nasty MDs. Section 2 presents a detailed investigation of the malicious attacks in MANET and as prevention works as the first line of defense, a detail survey of various selfish behaviour prevention schemes is presented. Finally the discussion is concluded in Section 3.

## 2 RELATED WORK

With growing popularity of MANET, it is becoming critically important to provide similar QoS to the users as they are accustomed to in networks with wired infrastructure. The most interesting difference lies in the fact that links are relatively unreliable, dynamic, and resource constraint in nature. Moreover, the unprotected locations of wireless routers expose them to malicious intrusions and environmental hazards, such as, thunder storms, lightening, etc., adding extra vulnerability in network operations. As a consequence of faults (natural and man-made), MANET might perform inefficiently and could be temporarily shut down in extreme cases. Limited resources and shared bandwidth lead to degraded per-

————————————————

*Dr. P.K.Suri is Professor with the Deptt. Of Comp. Science & Applications, Kurushetra University, Kurukshetra, Haryana, India.*
*Mrs. Kavita Taneja is Assistant Proessor with M.M. Inst. of Comp. Tech. & B. Mgmt., Maharishi Markandeshwar University, Mullana, Haryana, India.*

formance in MANET comparing with wired systems. However, some resources are consumed quickly as the MUs participate in the network functions. Battery power is considered to be most important in a mobile environment. One of the major sources of energy consumption in the MUs of MANETs is wireless transmission. An individual MU may attempt to benefit from other MUs, but refuse to share its own resources. Such MUs are called selfish or misbehaving units and their behavior is termed as misbehavior [6]. A selfish unit may refuse to forward data packets for other MUs in order to conserve its own power.

## 2.1 Misbehaviour in MANET

The main assumption of the MANET routing protocols is that all participating MDs do so in good faith and without maliciously disrupting the operation of the protocol [7]. However, the existence of malicious units cannot be overlooked in any system, especially in open ones like social MANET. Our analysis focuses only on network-layer threats and countermeasures and depth analysis brings into light new parameters of distrust and extends new definition of selfish behaviour. Nasty MDs have complete access to the communication link they are able to advertise false routing information at will and force arbitrary routing decisions on their peers [8]. Based on the threat analysis and the identified capabilities of the potential attackers, we will now discuss several specific attacks that can target the operation of a routing protocol in MANET.

*Location disclosure [9]:* Location disclosure is an attack that targets the privacy requirements of MANET. Through the use of traffic analysis techniques [10] or with simpler probing and monitoring approaches an attacker is able to discover the location of a MD, or even the structure of the entire network.

*Black hole [8]:* In a black hole attack a nasty MD injects false route replies to the route requests it receives advertising itself as having the shortest path to a destination. These fake replies can be fabricated to divert network traffic through the malicious node for eavesdropping, or simply to attract all traffic to it in order to perform a denial of service attack by dropping the received packets.

*Wormhole [11]:* The wormhole attack is one of the most powerful presented here since it involves the cooperation between two malicious nodes that participate in the network. One attacker, say node A, captures routing traffic at one point of the network and tunnels them to another point in the network, say to node B, that shares a private communication link with A. Node B then selectively injects tunneled traffic back into the network. The connectivity of the nodes that have established routes over the wormhole link is completely under the control of the two colluding attackers.

*Blackmail [12]:* This attack is relevant against routing protocols that use mechanisms for the identification of malicious nodes and propagate messages that try to blacklist the offender. An attacker may fabricate such reporting messages and try to isolate legitimate nodes from the network. The security property of nonrepudiation can prove to be useful in such cases since it binds a MD to the messages it generated [13].

*Denial of service:* Denial of service attacks aim at the complete disruption of the routing function and therefore the whole operation of the adhoc network. Specific instances of denial of service attacks include the routing table overflow [9] and the sleep deprivation torture [14]. In a routing table overflow attack the malicious MD floods the network with bogus route creation packets in order to consume the resources of the participating MDs and disrupt the establishment of legitimate routes. The sleep deprivation torture aims at the consumption of batteries of a specific MD by constantly keeping it engaged in routing decisions.

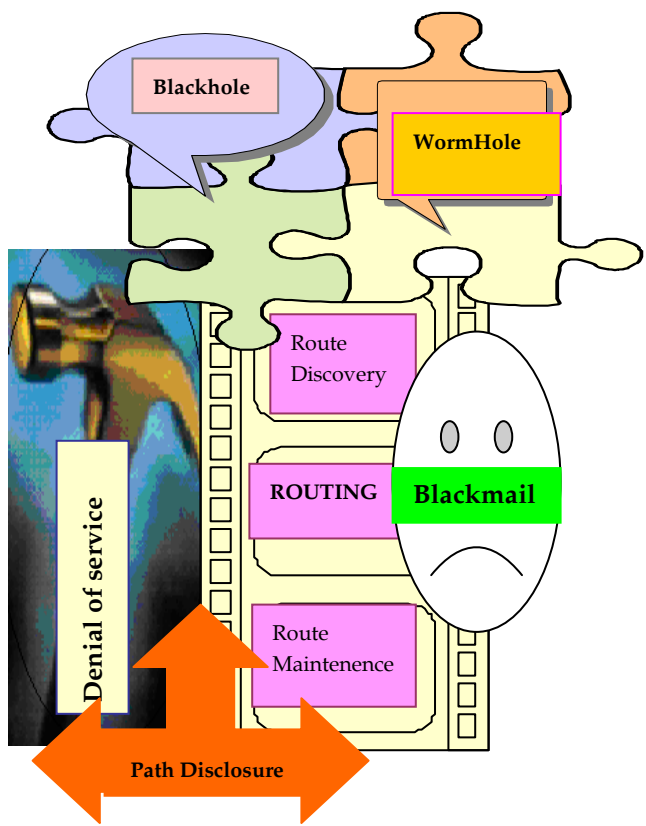

Figure1. Misbehaviour in MANET

## 2.2 Preventing Selfishness

The misbehavior problem of certain MUs in MANETs has lead to techniques to prevent selfishness in MANETs, and this paper presents detailed analysis of the same as classification into three categories reputation-based schemes, credit-based schemes and game theory based schemes (Table 1-3). Reputation is the opinion of the public toward a person, a group of people, an organization, or a resource. In the context of collaborative applications such as peer-to-peer systems, reputation represents the opinions MDs in the system have about their peers and peer-provided resources. Reputation allows parties to build confidence, or the degree to which one party trusts another within the context of a given purpose or decision. By harnessing the community knowledge in the form of feedback, reputation-based systems help participants decide whom to trust, encourage trustworthy behavior, and discourage dishonest participation by providing a means through which reputation and ultimately trust can be quantified and disseminated. Reputation-Based Scheme [15-20] works on the theme of detection of misbehaving MUs and their boycott from any communication. Each MU participates equally in the absence of any central administration to collectively detect and declare the misbehavior of a suspicious MU. Such a declaration is then broadcasted throughout the network so that the selfish MU will be cut off from the rest of the network. Reputation-Based Scheme [17] contains two major modules, termed watchdog and pathrater, to detect and mitigate, respectively, routing misbehavior in MANETs. MUs operate in a flexible mode wherein the watchdog module overhears the medium to check whether the next-hop MD faithfully forwards the packet. Based on the watchdog's labels, the path rater module rates every path and subsequently selects the path that best avoids misbehaving MDs. The drawback is complete reliance on overhearing of a medium that is already mysterious of noise and errors, so it has been found that the watchdog technique fail to detect misbehavior or raise false alarms in the presence of ambiguous collisions, receiver collisions, and limited transmission power. The CONFIDANT protocol [17] is another example of reputation-based schemes. The protocol is based on selective altruism and utilitarianism, thus making misbehavior unattractive. CONFIDANT consists of four important components—the Monitor, the Reputation System, the Path Manager, and the Trust Manager performing the neighborhood watching, MD rating, path rating, and sending and receiving alarm messages, overhearing technique. The scheme inherits the demerits of the watchdog scheme. The basic idea of credit-based schemes [21-25] is "serve and earn." Faithful MUs get incentives in terms of virtual currency for providing services to other MUs for efficient communication. When such MUs request other MUs for packet forwarding, the same payment system to pay for such services is implemented. The main hurdle in credit-based schemes is the need of extra protection for the virtual currency and/ or tamper-resistant hardware. Sprite [23] eliminates the use of tamper resistant hardware to stimulate cooperation among the selfish MDs. The need to have a central authority and falling short to consider malicious MD's packet drop causes Sprite to be a non-generic proposal. Game theory based [26,27] scheme follow Nash equilibrium including two kinds of games namely Cooperative game in which players reach an agreement through communication and Non-cooperative game in which players chase their own profit independently. GIFT (Generous TIT-FOR-TAT) each MU keeps track of the ratio of services provided to services taken. But it requires additional information per session leading to overhead.

## 3. CONCLUSION AND FUTURE WORK

MANETs are particularly sensible to unexpected behaviors. The lack of a common goal in open MANETs without a centralized authority will make their maintenance tough: each user will attempt to retrieve the most of the network while expecting to pay as less as possible. In human communities, this kind of behavior is called selfishness. We analyzed the different types of attacks in an ad-hoc environment. While prohibiting selfishness shows to be impossible over a decentralized network, applying punishments to those that present this behavior may be beneficial. This position paper details the analysis of selfishness prevention schemes for MANET.

TABLE 1 REPUTATION BASED SELFISHNESS PREVENTION SCHEMES

| Selfish MD prevention scheme | Components | Checkpoints and limitations. |
|---|---|---|
| Watchdog & Pathrater | ➤ **Watchdog:** Overhears neighbors' packet transmission promiscuously notifies misbehavior to the source MS by sending a message.<br>➤ **Pathrater:** Collects notification and rates every other MS avoids unreliable MSs in finding a path. | ✓ Depends only on Promiscuous listening.<br>✓ Selfish MDs are just bypassed and not punished.<br>✓ Promiscuous mode doesn't work with unidirectional link or directional antenna. |
| CONFIDANT | ➤ **Monitor:** promiscuously observes route protocol behavior as well as packet transmission of neighbor MDs.<br>➤ **Trust manager:** sends ALARM messages on detection of misbehavior.<br>➤ **Reputation system:** maintains a rating list and a blacklist for other MDs.<br>➤ **Path manager:** ranks paths according to the reputation of MDs along each path. | ✓ Uses both direct and indirect observations from other MDs.<br>✓ Similar weakness as Watchdog & Pathrater. |
| CORE (COllaborative Reputation) | ➢ **Subjective Reputation:** own observation on usual packet transmission<br>➢ **Functional Reputation:** own observation on task specific behavior of neighbor MDs<br>➢ **Indirect Reputation:** positive reports from other MDs<br>➢ **False Accusation:** by malicious MDs is prevented | ✓ MDs with bad reputation are isolated.<br>✓ Aging factor with more weight on past observation hence slow reaction.<br>✓ Reputation of MDs is not changed frequently, thus MDs temporarily suffering from bad environmental conditions are not punished severely. |
| Friends & Foes | ➢ **Friends:** MDs to which MD is willing to provide service.<br>➢ **Foes:** MDs for which MD refuses to serve.<br>➢ **Selfish:** MDs that regard MD as a foe.<br>➢ **Friends -> Foes:** When the difference between the numbers of packets forwarded for each other, reach a threshold value.<br>➢ **Friends -> Selfish:** When a neighbor does not fairly treated packets forwarded by MD A (promiscuous listening). | ✓ Memory overhead.<br>✓ Message overhead. |
| Reputation Indexing Window (RIW) | ➢ Contrary to CORE, emphasis on **current feedback items** rather than old ones.<br>➢ Keeps a MD from behaving selfishly for a long time after building up good reputation. | ✓ Arbitrary weight without theoretic base. |

**TABLE 2** CREDIT BASED SELFISHNESS PREVENTION SCHEMES

| Selfish MD Prevention Scheme | Components | Checkpoints and Limitations. |
|---|---|---|
| PPM / PTM | ➢ **Packet Purse Model:** Source loads Nuglets into data packets before sending them for the payment to intermediate MDs Intermediate MDs can take more Nuglets than they deserve.<br>➢ **Packet Trade Model:** Intermediate MDs trade packets with the previous MD, and the destination finally pays the price of the packets Easy target of the Denial-of-Service (DoS) attack. | ✓ Require a secure hardware to keep MDs from tampering the amount of Nuglets. |
| Sprite<br>Simple, cheat-proof, credit based system | ➢ **CCS (Central Authorized Server)**- MDs send CCS a receipt for every packet they forward, CCS gives credits to MDs according to the receipt. | ✓ Scalibility is major obstacle.<br>✓ Message overhead |
| Ad hoc-VCG(Vickery, Clarke and Groves) | ➢ **Route Discovery**- Destination MD computes needed payments for intermediate MDs and notifies it to the source MD or the central bank.<br>➢ **Data Transmission**- Actual payment is performed in this phase. | ✓ Totally depends on destination MD's report. |
| PIFA (Protocol independent Fairness Algorithm) | ➢ **Credit Manager (CM):** manages the credit database.<br>➢ **MD**: who has the maximum energy<br>➢ **Base station** or sink MD | ✓ Compatible to any routing protocol.<br>✓ Message processing overhead in CM |

**TABLE 3** GAME THEORY BASED SELFISHNESS PREVENTION SCHEMES

| Selfish MD prevention scheme | Components | Checkpoints and limitations. |
|---|---|---|
| GIFT (Generous Tit For Tat) | ➢ Pursues energy consumption fairness for a session as a unit low complexity.<br>➢ For every session, MDs try to balance between service provided to them and service provided by them. | ✓ Help received + delta > Help given Each MD needs sufficient information about an entire system, e.g., the number of MDs, energy constraints of them, and request for each session |
| CATCH | ➢ **Anonymous Challenge Message (ACM):** Anonymous ACM messages with sender ID hidden | ✓ MD should have at least one connection to send its own packet, so it has no choice but acknowledge |
| | are sent by tester to check connectivity of testee.<br>➢ **Anonymous Neighbor Verification (ANV):** In the ANV phase, a tester sends a cryptographic hash of a random token and releases the secret token to it if testee is good. | all anonymous messages<br>✓ No proof for evolutionary stability |
| SLAC | ➢ Compares its performance against other MDs Based on a repeated game.<br>➢ Prisoner's dilemma in P2P networks. | ✓ Fair comparison of MD performance required |
| Incentive Scheduling | ➢ Centrally managed and provide more time slots and power for relay MDs than non relay MDs | ✓ Relay MD actually not relaying packets |